\documentclass[10pt]{article}
\usepackage{amssymb}
\usepackage{amscd}
\usepackage{latexsym}
\usepackage{endnotes}
\usepackage{graphicx}
\newcommand{\Rr}{{\mathbb R}}

\def\be{\begin{equation}}
\def\ee{\end{equation}}
\def\bea{\begin{eqnarray}}
\def\eea{\end{eqnarray}}

\def\d{{\,\rm d}}

\def\0{{\bf 0}}

\def\p{{\bf p}}

\def\P{{\bf P}}

\def\h2m{\frac{\hbar^2}{2m}}
\def\p0{{P_{\beta H^0_N}}}

\newtheorem{theorem}{Theorem}

\begin{document}
\title{{\flushleft{\small {\rm Published in J. Phys. A: Math. Theor. \textbf{44}, 035205 (2011)}\\}}\vspace{1cm}
\large\bf Superimposed particles in 1D ground states}
\author{Andr\'as S\"ut\H o\\Research Institute for Solid State Physics and Optics\\Hungarian Academy of Sciences\\P. O. B. 49, H-1525 Budapest, Hungary\\
E-mail: suto@szfki.hu\\}
\date{}
\maketitle
\thispagestyle{empty}
\begin{abstract}
\noindent
For a class of nonnegative, range-1 pair potentials in one dimensional continuous space we prove that any classical ground state of lower density $\geq 1$ is a tower-lattice, i.e., a lattice formed by towers of particles the heights of which can differ only by one, and the lattice constant is 1. The potential may be flat or may have a cusp at the origin; it can be continuous, but its derivative has a jump at 1. The result is valid on finite intervals or rings of integer length and on the whole line.

\vspace{2mm}
\noindent
PACS: 61.50.Ah, 02.30.Nw, 61.50.Lt

\vspace{2mm}
\end{abstract}

\section{Introduction}
A curious aspect of soft potentials is that they allow classical particles to superimpose in ground state configurations (GSCs), even if the pair potential is purely repulsive~\cite{Lik}. This property is related to the Fourier transform of the interaction and can appear | mostly but not exclusively | if this one is partially negative. Indeed, recently it was proven that a strictly positive Fourier transform leads to a uniform distribution of particles at high densities, while a partially negative Fourier transform causes non-uniformity~\cite{Suto}. However, the forms of non-uniformity can be various, and the precise conditions on the Fourier transform giving rise to superimposed particles are still unknown. The three-dimensional example of Likos et al.~\cite{Lik} is the pair potential $e^{-\alpha|x|^m}$ with $m>2$. In the case $m=4$ a detailed numerical analysis and analytic arguments show rather convincingly the superposition of particles. The second derivatives of these interactions vanish at the origin. This implies a strong negative part of the Fourier transform at large wave vectors, and intuition also suggests that the flatness of the interaction at zero distance favors the formation of towers of particles at lattice sites. The type of the lattice is less intuitive. If the (negative) minimum of the Fourier transform decides the lattice type, the reciprocal lattice should be close-packed. Until now a rigorous proof exists only in the case of the penetrable-sphere model, particles with a repulsive square core potential~\cite{Suto}. This is the flattest possible interaction, its Fourier transform is oscillating and slowly decaying, and not the reciprocal but the real-space lattice is close-packed, the lattice constant being the range of the potential. Mathematically, this problem is almost trivial and, because of the jump of the interaction, the model is pathological in the sense that the energy and the free energy do not tend to the energy of the GSCs as the temperature goes to zero~\cite{Suto}. Below we present a one-dimensional generalization of the penetrable-sphere model, providing less trivial examples. We shall also see that a vanishing second derivative of the interaction at zero distance is not a prerequisite for the phenomenon to occur (this was noted already in \cite{Lik}). The extreme value minus infinity for the second derivative, i.e., a cusp at the origin is also allowed. This is particularly interesting in comparison with other examples of interactions which have a cusp at zero and a partially negative Fourier transform, and no Bravais lattice, with or without superposition of particles, as a high-density GSC~\cite{Suto}.

\section{Theorem and proof}
Given an even translation invariant pair potential $u$, the potential energy of $N$ pointlike particles in the configuration $(x)_N:=(x_1,x_2,\ldots,x_N)$ is
\be
U(x)_N=\sum_{1\leq i<j\leq N}u(x_i-x_j).
\ee
Further on, $\{x\}^m$ will denote the sequence $x,\ldots,x$ where $x$ is repeated $m$ times. Recall that an $N$-particle GSC in a bounded domain $\Lambda$ is a configuration that minimizes $U(x)_N$ among $(x)_N\in\Lambda^N$, and a GSC in infinite space is an infinite configuration whose energy cannot be decreased by any number-preserving local perturbation~\cite{Suto}.

\begin{theorem}
In one dimension, let $u$ be a bounded real function with
\be\label{potential}
u(x)=u(-x)\left\{\begin{array}{ll}
=1,& x=0\\
\geq 1-|x|,& 0<|x|<1\\
=0, & |x|\geq 1.
\end{array}\right.
\ee
(i) GSCs on an interval of integer length.
\begin{enumerate}
\item
For arbitrary positive integers $n,m$, any $r\in\{0,1,\ldots,n\}$ and any integers $0\leq i_1<\cdots<i_r\leq n$ the configurations
\be\label{GS}
X^{n,m}_{i_1\ldots i_r} =\left(\{i_1\}^{m+1},\ldots,\{i_r\}^{m+1},\left(\{j\}^m\right)_{j\in \{0,\ldots,n\}\setminus\{i_1,\ldots,i_r\}}\right)
\ee
are $N=m(n+1)+r$-particle GSCs of $u$ on the interval $[0,n]$. That is, each integer point in $[0,n]$ is occupied by $m$ or $m+1$ particles. The energy of the $N$-particle GSCs is
\be\label{energy}
E_n(N)=(n+1){m\choose 2}+r m=U(X^{n,m}_{i_1\ldots i_r}).
\ee

\item
If $r=0$, the unique $N$-particle GSC in $[0,n]$ is $X^{n,m}=\left(\{j\}^m\right)_{j=0}^n$.
\item
If $u(x)>1-|x|$ for $0<|x|<1$ then $X^{n,m}_{i_1\ldots i_r}$ are the only N-particle GSCs in $[0,n]$ for every $r$.
\item
If $u$ is the `overlap potential', $u(x)=(1-|x|)1_{[-1,1]}(x)$, and $r>0$ then the ground state is continuously degenerate. On the background of $X^{n,m}$ the remaining $r$ particles can be freely distributed under the constraint that the distance of any two of them is $\geq 1$.
\end{enumerate}

\noindent
(ii) GSCs on the whole line.
\begin{enumerate}
\item
Any infinite configuration $X=\left(\{i\}^{m_i}\right)_{i\,=-\infty}^\infty$ with $m_i\geq 1$ and $|m_i-m_j|\leq 1$ for all $i,j$ is a GSC of $u$ on $\Rr$.
\item
The ground state of the overlap potential is continuously degenerate. For any integer $m\geq 1$, any $0\leq x_1\leq\ldots\leq x_m<1$ and any $Y=(\cdots<y_{-1}<y_0<y_1<\cdots)$ such that $y_{k+1}-y_k\geq 1$ for all $k$,
\be\label{GSC-overlap}
X=\bigcup_{j=1}^m\bigcup_{i=-\infty}^\infty \{i+x_j\}\quad\mbox{and}\quad X\cup Y
\ee
are GSCs. Here the union is in the sense of superposition, i.e. coinciding points are multiply counted.

\item
If a GSC has a density, and the value of this is $\rho$, then the ground-state energy per unit length is
\be\label{energy-dens}
\lim_{n\to\infty}\frac{E_n(N)}{n}=\frac{1}{2}\lfloor\rho\rfloor(\rho+\{\rho\}-1)
\ee
where $\lfloor\rho\rfloor$ and $\{\rho\}$ are the integer and fractional parts of $\rho$, respectively.
\end{enumerate}

\noindent
(iii) GSCs on a ring of integer length.

\noindent
For $n\geq 1$ let $\Lambda=[0,n+1[$ and $u_\Lambda(x)=\sum_{j=-\infty}^\infty u(x+j(n+1))$. Any $N\geq n+1$-particle GSC of $u$ on $[0,n]$ is a GSC of $u_\Lambda$ on $\Lambda$ with energy $E_{n}(N)$. In the case of the overlap potential the restriction to $[0,n+1[$ of any GSC of the type (\ref{GSC-overlap}) is also a GSC of $u_\Lambda$.
\end{theorem}

\noindent
\emph{Proof.} (i/1). Let $0\leq x_1\leq x_2\leq\cdots\leq x_N\leq n$ be any $N$-particle configuration:
\be\label{U-lower}
U(x)_N=\sum_{1\leq i<j\leq N}u(x_i-x_j)\geq \sum_{i\geq 1}\sum_{j=1}^mu(x_i-x_{i+j}).
\ee
If $(x)_N=X^{n,m}_{i_1\ldots i_r}$ then (\ref{U-lower}) holds with equality because $x_{i+j}-x_i\geq 1$ for $j\geq m+1$. The number of retained pairs is \be\label{|pairs|}
\#\{(i,i+j)|i\geq 1, 1\leq j\leq m, i+j\leq N\}=N-1+N-2+\cdots+N-m=mN-{m+1\choose 2}.
\ee
The lower bound (\ref{U-lower}) can be rearranged into a summation over chains of pairs,
\be\label{chain}
\sum_{i\geq 1}\sum_{j=1}^mu(x_i-x_{i+j})=\sum_{j=1}^m\sum_{i=1}^j [u(x_i-x_{i+j})+u(x_{i+j}-x_{i+2j})+u(x_{i+2j}-x_{i+3j})+\cdots].
\ee
There are altogether $\sum_{j=1}^m j={m+1\choose 2}$ chains. Now we apply the lower bound
$$u(x-y)\geq 1+x-y,$$
valid for any pair $x\leq y$, to each term of (\ref{chain}). The sum of the unities gives the total number of pairs (\ref{|pairs|}). The sum over the differences of particle coordinates in a chain gives the smallest minus the largest coordinate, and can be bounded from below by $-n$. Thus,
\bea\label{r-general}
U(x)_N\geq m N-{m+1\choose 2}-n{m+1\choose 2}=m[m(n+1)+r]-(n+1){m+1\choose 2}\nonumber\\
=(n+1){m\choose 2}+rm=U(X^{n,m}_{i_1\ldots i_r})
\eea
where the last equality is obvious. This shows that $X^{n,m}_{i_1\ldots i_r}$ is a GSC and proves the formula (\ref{energy}) for the ground-state energy.

(i/2). Let $r=0$, i.e. $N=m(n+1)$. If $X^{n,m}=(\xi)_N$, $\xi_i\leq\xi_{i+1}$, then
\be
U(X^{n,m})=\sum_{i\geq 1}\sum_{j=1}^{m-1}u(\xi_i-\xi_{i+j})
\ee
and, instead of (\ref{chain}), it suffices to sum over $j$ up to $m-1$. In
$$\{(i,i+j)|i\geq 1, 1\leq j\leq m-1, i+j\leq N\}$$
the number of pairs is $(m-1)N-{m\choose 2}$ and the number of chains is ${m\choose 2}$. This yields the lower bound
\be
\sum_{i\geq 1}\sum_{j=1}^{m-1}u(x_i-x_{i+j})\geq (m-1)N-(n+1){m\choose 2}=(m-1)m(n+1)-(n+1){m\choose 2}=(n+1){m\choose 2}=E_n(N).
\ee
We show that if $(x)_N\neq X^{n,m}$, then there is a pair $(k,k+m)$ such that $x_{k+m}-x_k<1$, and therefore
\be
U(x)_N\geq \sum_{i\geq 1}\sum_{j=1}^{m-1}u(x_i-x_{i+j})+u(x_k-x_{k+m})\geq E_n(N)+1+x_k-x_{k+m}>E_n(N).
\ee
Assume that $x_{k+m}-x_k\geq 1$ for all $k$. Then
\be
n\geq \sum_{i=1}^n (x_{im+k}-x_{(i-1)m+k})=x_{nm+k}-x_k\geq n
\ee
for $k=1,2,\ldots,m$. Thus,
$$x_1=x_2=\cdots=x_m=0\quad\mbox{and}\quad x_{nm+1}=x_{nm+2}=\cdots=x_{(n+1)m}=n.$$
It follows that $x_{m+1}\geq 1$ and $x_{nm}\leq n-1$. Then
\be
n-2\geq \sum_{i=2}^{n-1} (x_{im+k}-x_{(i-1)m+k})=x_{(n-1)m+k}-x_{m+k}\geq n-2
\ee
for $k=1,2,\ldots,m$. Thus,
$$x_{m+1}=x_{m+2}=\cdots=x_{2m}=1\quad \mbox{and}\quad x_{(n-1)m+1}=x_{(n-1)m+2}=\cdots=x_{nm}=n-1$$
and it follows that $x_{2m+1}\geq 2$ and $x_{(n-1)m}\leq n-2$. Further repeating this argument, in $\lfloor n/2\rfloor$ steps we find that the unique $(x)_N$ in $[0,n]$ satisfying $x_{k+m}-x_k\geq 1$ for all $k$ is $X^{n,m}$.

(i/3). Consider the case when $u(x)>1-|x|$ for $0<|x|<1$. We may suppose $r>0$. Let $(x)_N$ be such that $0<x_k-x_i<1$ for some pair $i<k$. If $k>i+m$, the positive contribution of $u(x_i-x_k)$ was not taken into account in (\ref{U-lower}). If $k\leq i+m$, $u(x_i-x_k)$ was taken into account with the lower bound $1+x_i-x_k$. In either case we obtain $U(x)_N>E_n(N)$. Thus, in any GSC, for any $i<k$ we have either $x_k-x_i\geq 1$ or $x_i=x_k$ with $k\leq i+m$. Necessarily, $x_{k+m+1}-x_k\geq 1$ for all $k$. As in (i/2), one can then prove that $X^{n,m}_{i_1\ldots i_r}$ are the only configurations satisfying this property.

(i/4). If $u(x)=1-|x|$ for $|x|\leq 1$ and is zero otherwise then
\be\label{overlap-degen}
U(\{i\}^m,\{x\},\{i+1\}^m)=m^2,
\ee
whenever $x\in[i,i+1]$, and the particle in $x$ does not interact with particles not in $i$ or $i+1$. This proves the assertion.

(ii/1). Take any $X=\left(\{i\}^{m_i}\right)_{i\,=-\infty}^\infty$ with $m_i\geq 1$ and $|m_i-m_j|\leq 1$ for all $i,j$. Thus, $m_i$ can assume only two values, say, $m(\geq 1)$ and $m+1$. Any local modification of $X$ is confined in an interval $[n_1,n_2]$ where $n_1<n_2$ are integers. According to (i), $X\cap[n_1,n_2]$ is a GSC in $[n_1,n_2]$, meaning that for any $N=|X\cap[n_1,n_2]|$-particle configuration $(y)_N\subset [n_1,n_2]$,
\be\label{inside}
U(y)_N\geq U(X\cap[n_1,n_2]).
\ee
Because the distance of both $(y)_N$ and $X\cap [n_1,n_2]$ to the rest of $X$ is at least 1, there is no interaction between particles inside $[n_1,n_2]$ and in $X\setminus [n_1,n_2]$. Together with (\ref{inside}) this implies that $X$ is locally stable, that is, a GSC on $\Rr$~\cite{Suto}.

(ii/2). This could be derived from Eq.~(\ref{overlap-degen}), but there is a more transparent geometric proof: $u(x)=(1-|x|)1_{[-1,1]}(x)$ is called the overlap potential because it measures the overlap of two aligned rods of unit length such that their centers are at a distance $|x|$. Replace each particle of $X$ or of $X\cup Y$ by a rod (interval) of unit length centered at that particle. Then no local displacement of the rods can decrease the sum of their overlaps, showing that $X$ and $X\cup Y$ are locally stable.

(ii/3). For sufficiently fast decaying interactions the energy density, if exists, is the same (and minimal) for all locally stable configurations of the same density. This \emph{absence of metastability} was proven in~\cite{SuPRB}. We can therefore compute the energy density from any GSC of a given density. Choose $X=\left(\{i\}^{m_i}\right)_{i\,=-\infty}^\infty$ with $|m_i-m_j|\leq 1$. Let $I$ and $\ell(I)$ denote a finite interval and its length, respectively. The lower and upper densities of $X$ are defined as
\be
\underline{\rho}(X)=\liminf_{\ell(I)\to\infty}\frac{|X\cap I|}{\ell(I)},\quad \overline{\rho}(X)= \limsup_{\ell(I)\to\infty}\frac{|X\cap I|}{\ell(I)}
\ee
where the limit is taken on arbitrary sequences of intervals. Suppose that $$\underline{\rho}(X)=\overline{\rho}(X)=\rho\in [m,m+1[.$$
Let $I=[0,n]$ and $N=|X\cap [0,n]|$. If $n$ is large enough then $N=m(n+1)+r$ where $0\leq r\leq n$ and $r$ depends on $I$. However, because of the existence of the limit defining the density, the limit of $E_n(N)/n$ also exists and equals (\ref{energy-dens}).

(iii) This is the case of a periodic boundary condition, and $u_\Lambda$ is the periodized interaction. Because the range of $u$ is 1, in $u_\Lambda$ only a single term can be nonzero. The difference and relative subtlety compared with case (i) comes from the fact that now the length of the interval is larger by 1 for the same number of particles, but the interaction between particles at the two ends of the interval adds to the energy. Here we refer to a more general result. Fix $N\geq n+1$ and let $X_0$ be one of the $N$-particle GSCs of $u$ on $[0,n]$. Let $X$ be the periodic extension of $X_0$ to $\Rr$. By part (ii), $X$ is a GSC of $u$ on $\Rr$. Then, according to Lemma 7.2 of Ref.~\cite{Suto}, $X\cap\Lambda$=$X_0$ is a GSC of $u_\Lambda$ on $\Lambda$. Furthermore,
\be
U_\Lambda(X_0)=\sum_{(x,y)\subset X_0}u_\Lambda(x-y)=U(X_0)=E_n(N). \qquad
\ee
In the case of the overlap potential we can directly restrict the configurations (\ref{GSC-overlap}) to $\Lambda$ and find the result for example by the geometric argument of (ii/2).

\section{Discussion}
The aim of this paper was to provide provable nontrivial examples of pair interactions which give rise to the superposition of particles in classical ground state configurations. Obviously, such an interaction must be bounded for touching particles. Trivial examples include unstable interactions, for examples, attractive ones. An almost trivial example is the penetrable sphere model. The important discovery of Likos et al.~\cite{Lik} was that the interaction can be repulsive everywhere outside the origin, what really counts is that it must have a partially negative Fourier transform. This led them to the study of the family $e^{-\alpha|x|^m}$ with $m>2$. Although the superposition has been established numerically, a mathematical proof is still missing for this class of potentials. On the other hand, in~\cite{Suto} counterexamples were constructed, showing that the partial negativity of the Fourier transform was not sufficient | it was, however, proven to be `almost necessary' in the sense that a strictly positive Fourier transform causes the particles to distribute uniformly as the density increases. The examples given in the present paper are nonnegative range-1 potentials which may or may not be purely repulsive within their range. The GSCs at densities $\rho<1$ are trivial and form a continuously degenerate family (any configuration in which the distance of neighboring particles is larger than or equal to 1 is a GSC), while in GSCs for $\rho\geq 1$ the particles superimpose on the sites of a lattice of lattice constant 1. It is quite possible, however, that for `generic' interactions having a partially negative Fourier transform the high-density GSCs are not tower-lattices but periodic configurations in which the particles cluster \emph{around} (and not \emph{on}) the sites of a lattice.

Potentials whose Fourier transform is nonnegative and takes on zero somewhere represent a marginal case. They can be obtained as limits of potentials having a partially negative Fourier transform. If these latter have tower-lattice GSCs, the property extends to the limit by continuity, but the tower-lattice is expected to be embedded in a continuum of other GSCs. Here, this is the case of the overlap potential (previously studied by Torquato and Stillinger~\cite{TS}) whose Fourier transform is $(2/k^2)(1-\cos k)$. At integer densities the ground state of the overlap potential on an interval of integer length is non-degenerate, but in infinite space and on the ring we meet the same kind of continuous degeneracy as in the case of pair potentials with a nonnegative Fourier transform of compact support~\cite{Su2}.

There has been no use of Fourier transform in our proof. This may seem curious in view of the decisive role of the Fourier transform of the interaction. However, in $k$-space a proof would have been less complete. Based on Proposition 4.2 of Ref.~\cite{Suto}, for integer densities $\rho=m$ (and only for them) we can prove that $X^{n,m}$ are GSCs of $u_\Lambda$ on $\Lambda=[0,n+1[$ by showing that among positive measures $\mu$ of total weight $n+1$ on $\Lambda$ the functional
\be
I[\mu]=\frac{1}{2(n+1)}\int_0^{n+1}\int_0^{n+1}u_\Lambda(x-y)\mu(\d x)\mu(\d y)
\ee
is minimized by $\mu_0=\sum_{j=0}^n\delta_j$ (here $\delta_j$ is the Dirac delta localized at $j$). In Fourier representation $I[\mu]$ reads
\be
I[\mu]=\frac{1}{2}\sum_{j=-\infty}^\infty\widehat{u}\left(\frac{2j\pi}{n+1}\right) \left|\widehat{\mu}\left(\frac{2j\pi}{n+1}\right)\right|^2
\ee
where
\be
\widehat{u}(k)=\int u(x)e^{-ikx}\d x,\qquad \widehat{\mu}(k)=\frac{1}{n+1}\int_0^{n+1}e^{-ikx}\mu(\d x).
\ee
Therefore, one must prove that
\be
I[\mu_0]=\frac{1}{2}\sum_{l=-\infty}^\infty\widehat{u}(2l\pi)=\frac{u(0)}{2}=\frac{1}{2}\leq I[\mu].
\ee
For the overlap potential $I[\mu_0]$ is easily seen to be the minimum, and because the other potentials of the family (\ref{potential}) are bounded below by the overlap potential, $I[\mu_0]$ is the minimum of $I[\mu]$ also for them.

\vspace{5pt}
\noindent
\emph{Acknowledgement.} This work was partially supported by OTKA Grants K67980 and K77629.

\end{document}